\documentclass[aps,superscriptaddress,twocolumn,linenumbers, nofootinbib]{revtex4}
\usepackage{graphicx, amsfonts, amsmath, amssymb, physics, bm, amsthm}
\usepackage{xcolor}
\usepackage[hidelinks]{hyperref}
\usepackage{cleveref}
\usepackage{quantikz}
\usepackage{stmaryrd}

\newtheorem{theorem}{Theorem}
\newtheorem*{theorem*}{Theorem}
\newtheorem{lemma}{Lemma}
\newtheorem{definition}{Definition}
\newtheorem{corollary}{Corollary}

\renewcommand{\max}{\ensuremath{{\rm max}}}
\renewcommand{\min}{\ensuremath{{\rm min}}}
\newcommand{\lm}{\lambda_{\min}}
\newcommand{\lM}{\lambda_{\max}}

\newcommand{\aqa}{$\langle aQa ^L\rangle $ Applied Quantum Algorithms, Universiteit Leiden, the Netherlands}
\newcommand{\lorentz}{Instituut-Lorentz, Universiteit Leiden, the Netherlands}
\newcommand{\cern}{Quantum Technology Initiative, CERN, Geneva, Switzerland}

\usepackage{url}

\begin{document}

\title{Parameterized quantum circuits as universal generative models for continuous multivariate distributions}

\author{Alice Barthe}
\affiliation{\aqa}
\affiliation{\cern}
\affiliation{\lorentz}

\author{Michele Grossi}
\affiliation{\cern}

\author{Sofia Vallecorsa}
\affiliation{\cern}

\author{Jordi Tura}
\affiliation{\aqa}
\affiliation{\lorentz}

\author{Vedran Dunjko}
\affiliation{\aqa}
\affiliation{\lorentz}

\begin{abstract}
Parameterized quantum circuits have been extensively used as the basis for machine learning models in regression, classification, and generative tasks. 
For supervised learning, their expressivity has been thoroughly investigated and several universality properties have been proven.
However, in the case of quantum generative modelling, much less is known, especially when the task is to model distributions over continuous variables.
In this work, we elucidate expectation value sampling-based models.
Such models output the expectation values of a set of fixed observables from a quantum circuit into which classical random data has been uploaded.
We prove the universality of such variational quantum algorithms for the generation of multivariate distributions.
We explore various architectures which allow universality and prove tight bounds connecting the minimal required qubit number, and the minimal required number of measurements needed.
Our results may help guide the design of future quantum circuits in generative modelling tasks.
\end{abstract}
\maketitle

\section{Introduction}

Parameterized quantum circuits are the centrepiece of numerous approaches to machine learning on quantum computers, motivated by numerous near-term hardware limitations \cite{bharti_noisy_2022, cerezo_variational_2021}. The use of these models for supervised learning has been widely studied, for example in solving regression problems \cite{macaluso_variational_2020} where the goal is to assign continuous labels to data points. Variational algorithms have also been explored in so-called generative modelling tasks, where the objective is to generate new samples following a distribution that generated the training data \cite{liu_differentiable_2018}. 

A prominent example of distributions with discrete support is the quantum Born machine \cite{liu_differentiable_2018}, which stores a distribution over $n$-bit strings in a $n$-qubit state. Another model for discrete distributions is the quantum Boltzmann machine \cite{amin_quantum_2018}.
Going beyond distribution with discrete support, an approach has been introduced to model distributions where the random variable can in principle take on any value within a continuous interval. In such models, the quantum circuit takes classical randomness as input and outputs expectation values, consequently, we call this model expectation value samplers. This model has been extensively used as a quantum generator in the context of quantum generative adversarial networks \cite{romero_variational_2021,dallaire-demers_quantum_2018}.

While for quantum Born machines, the expressivity and universality have been clarified \cite{liu_differentiable_2018, coyle_born_2020}, in contrast, expectation value sampling models are not so well understood. In particular, an interesting feature of expectation value sampling is that the dimension of the output is not inherently tied to the number of qubits used. In this work, we focus on the expressivity of the generators based on expectation value sampling depending on the number of qubits and the spectrum of the observables. 

In the first section, we introduce and define expectation value sampling. We define formally what a universal generative model family is, and give some background on random variable transformation and some fundamental properties.

The second section addresses whether every arbitrary distribution can be modelled using expectation value sampling, i.e. whether this model is universal. We recall existing results on the universality of parameterized quantum circuits. We extend existing results and use them to constructively prove the existence of two families of universal expectation value models.

Following this, in the third section, using variants of the Holevo bound, we look into the necessary conditions for an expectation value scheme to be universal. We show that there exists a necessary condition as an upper bound on the dimensionality of the target distribution as a combination of the number of qubits and the observable spectral norm. By leveraging the scaling with the observable norm, we show that there exists a trade-off between the number of measurements and the number of qubits to reach universality. 

The final fourth section provides tools to analyse how the number of qubits, the data encoding scheme, and the set of observables affect the expressivity of the expectation value sampling models, with a focus on the dependence relationship between random variables. 

\section{Background and Definitions}
In this section, we give formal definitions of expectation value sampling models and universal generative model families. We also introduce random variable transformations, a core concept in this work.

\subsection{Definition of expectation value sampling}
\label{subsec:def_evs}
\begin{figure}
    \centering
    \begin{quantikz}[row sep=0.1cm, column sep=0.2cm]
        \lstick{$\ket{0}$} & \qw & \gate[3]{U_{\theta}(X)} & \qw & \meter[3]{}  \\
        \lstick{$\ket{0}$} &&&& \rstick{$Y = [\langle O_m \rangle ] \sim p_Y$ when $X \sim p_X$ }\\
        \lstick{$\ket{0}$} &&&& 
    \end{quantikz}
    \caption{Expectation value sampling model: a random vector is classically sampled. It is used to generate a random quantum state using a parameterized quantum circuit. The expectation values of fixed observables are returned as the output random vector.}
    \label{fig:EVScircuit}
\end{figure}
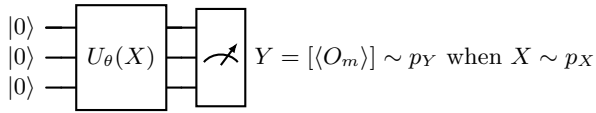

The expectation value sampling procedure goes as follows. A random variable is classically sampled and used as input to a parameterized quantum circuit which specifies a random quantum state. The expectation values of fixed observables are measured and returned as another random variable. We illustrate this procedure in \Cref{fig:EVScircuit}. It is important to note that, in contrast to the quantum circuit Born machine, the randomness does not come from the measurement process, but from the classical randomness provided as an input to the quantum circuit. Formally we define an expectation value sampling model as follows.
\begin{definition}[Expectation value sampling model]
    \label{def:EVS}
    An expectation value sampling model on $n$ qubits is defined by $(U_{\theta},\mathbf{O},p_X)$, where $U_{\theta}: \mathcal{X} \subseteq \mathbb{R}^L \rightarrow \mathcal{U}(2^n)$ is a parameterized quantum circuit taking data as input and returning a unitary matrix, $\mathbf{O} = (O_m)_{1 \leq m \leq M}$ is a vector of $M$ observables, and $p_X: \mathcal{X} \subseteq \mathbb{R}^L \rightarrow \mathbb{R}$ is the input distribution. We define the associated mapping $f$ as follows:
    \begin{equation}
        x \in \mathcal{X} \xrightarrow{f} \left(\bra{0} U_{\theta}(x)^\dag O_m U_{\theta}(x) \ket{0} \right)_{1 \leq m \leq M}.
    \end{equation}
    The output of the model is a sample drawn from the distribution $p_Y$ with $Y = f(X) \sim p_Y$ when $X \sim p_X$.
\end{definition}
In the above, the $N$-dimensional unitary group is denoted $\mathcal{U}(N)$.

Notably, unlike QCBMs, expectation value samplers have continuous support (absolutely continuous random variables, see \Cref{def:acrv}).

The central question of this work is whether such a model can generate any multivariate distribution, more precisely whether expectation value sampling is a universal generative model, according to the definition we will give in the next section.

\subsection{Definition of a universal generative model family}$
\label{subsec:univgenmod}$
In this subsection, we define universal generative model families.

\begin{definition}[Universal generative model family]
    \label{def:univgenmod}
    A generative model is a family of parameterized sampling procedures which enable the sampling from a corresponding set of $M$-dimensional probability density functions $\mathcal{P(\mathcal{X})}$ on $\mathcal{X} \subseteq \mathbb{R}^M$. 
    
    A generative model is called universal if for every probability density function $q$ on $\mathcal{X}$ there exists a sequence $\{p_k | p_k \in \mathcal{P(\mathcal{X})}\}_{1\leq k \leq \infty}$ such that the sequence of random variables $X_k \sim p_k$ converges in distribution to $ X \sim q$. 
    Equivalently, this means that the sequence of cumulative distribution functions of $p_k$, which we call $P_k$, converges pointwise to the cumulative distribution function of $q$, which we call $Q$.
    \begin{equation}
        \forall x \in \mathcal{X}, \lim_{k\rightarrow\infty} P_k(x)  = Q(x).
    \end{equation}
\end{definition}

This definition of closeness for random variables, also called convergence in distribution, is common in probability theory. For example, the central limit theorem precisely states the average of any $L$ independent random variables with mean $\mu$ and variance $\Sigma$ converges in distribution to the normal distribution $\mathcal{N}(\mu,\Sigma/L)$. 

In this work, we will mostly consider distributions with finite support, because this is the case in most practical real-world problems. In particular, their probability density functions are integrable functions and convergence in distribution implies convergence in the first Wasserstein distance $W_1$ \cite{villani_optimal_2009}. For completeness, we recall below the definition of the Wasserstein distance, also known as the Earth Mover's Distance, that we use in the context of this work.
\begin{definition}[Wasserstein distance]
    The $k$-th Wasserstein distance between two probability density functions $p$ and $q$ on $[-1,1]^M$ is defined as:
    \begin{equation}
    W_k(p, q) = \left( \inf_{\gamma \in \Pi(p, q)} \int_{\mathbb{R}^2} \|x - y\|^k d\gamma(x, y) \right)^{\frac{1}{k}},
    \end{equation}
    where $\Pi(p, q)$ is the set of couplings of $p$ and $q$, and $\| \cdot \|$ denotes the Euclidean distance. The parameter $k \geq 1$ determines the so-called order of the Wasserstein distance. 
\end{definition}
For the rest of the paper, we will only consider the first-order Wasserstein distance ($k=1$) and therefore simply refer to it as the Wasserstein distance.

\textbf{Note}: Universality is defined on a given support noted as $\mathcal{X}$ in \Cref{def:univgenmod}. For this work, we choose the support to be $\mathcal{X}=[-1,1]^M$, because the first step of most machine learning pipelines is to rescale the data to fit on a given interval.

Choosing universality on a cubic support $[-1,1]^M$ allows the expression of fully independent variables. Restricting $\mathcal{X}$ to smaller subsets would yield constraints on the dependence relationship expressible. For example, proving universality of a family of models for Dirichlet distributions would correspond to choosing $\mathcal{X} = \mathbb{S}_1^1$ where $\mathbb{S}_1^1\coloneqq \{x\in\mathbb{R}^M \geq 0 \mid \sum_m x_m = 1\}$ is the unit sphere for the $1$-norm.

\subsection{Random Variable Transformation}
A core concept in this work is that of random variable transformation. In this subsection, we introduce it and provide some of the associated fundamental properties. 

The first step in our analysis is the observation that an expectation value sampling model is a process that maps an input random vector (parameterizing the quantum circuit) to an output random vector (the expectation values of a set of observables). The literature on optimal transport and measure theory \cite{bogachev_triangular_2005} states that for every pair of absolutely continuous random variables of the same dimension, there exists a mapping to transform one into the other.
\begin{lemma}
     For every pair of probability density functions $p_X\in \mathcal{P}(\mathcal{X\subseteq \mathbb{R}^M})$ and $ p_Y\in \mathcal{P}(\mathcal{Y\subseteq \mathbb{R}^M})$, there exists a mapping $f : \mathcal{X} \rightarrow \mathcal{Y}$, that maps $Y\sim p_Y$ to $X\sim p_X$ as $Y=f(X)$.
\end{lemma}

We give intuition on how to construct this mapping and show that it can be chosen to be bounded piece-wise continuous in \Cref{app:triangular}.

Many generative modelling systems including Generative Adversarial Networks, Variational Auto-Encoders and normalizing flows rely on this idea to generate arbitrary distributions. In particular, we choose the initial distribution to be something simple, and then by altering the mapping applied to this random input, we obtain a rich spectrum of possible output distributions. 

 Then, sufficient conditions for a family of mappings to yield a universal generative model, in the sense of \Cref{def:univgenmod}, are well-known, and expressed in terms of the universality of mappings themselves. From \cite{kobyzev_normalizing_2021}, it is sufficient for a family of mappings to be dense in the set of all monotonically increasing functions in the pointwise convergence topology to yield a universal generative model. We explain the difference between pointwise topology and uniform topology in \Cref{app:uniformtopointwise}.
Since these are sufficient conditions and monotonic functions are included in all functions, the following holds.
\begin{theorem}
\label{thm:universalityflow}
    If a family of mappings $\mathcal{G} = \{g: \mathcal{X} \subseteq \mathbb{R}^M \rightarrow \mathcal{Y} \subseteq \mathbb{R}^M \}$ is dense in the set of all functions in the pointwise convergence topology, then this family of mappings $\mathcal{G}$ together with a probability density function $p_X$ with non zero support on $\mathcal{X}$ yields a universal generative model family on $\mathcal{Y}$ (cf \Cref{def:univgenmod}).
\end{theorem}

\textbf{Note}: In conventional machine learning, this notion of universality is common. It has been proven for several families of mappings in the context of normalizing flows, which are mappings with the additional property of being invertible: generic triangular mappings \cite{bogachev_triangular_2005}, neural networks mappings \cite{huang_neural_2018} and polynomial mappings \cite{jaini_sum_2019}.

In the next section, to prove the universality of expectation value samplers, we make explicit the connection between universal mapping families and the known results on the universality of parameterized quantum circuits as supervised learning models.

\section{Parameterized Quantum Circuits as Universal Generators}
\label{sec:sufficient}

\subsection{Universality of the product encoding circuit}
\label{subsec:univ_pe}
In recent literature, a number of universality properties of parameterized quantum circuits have been proven. In \cite{perez-salinas_one_2021}, a family of single qubit quantum circuits with an increasing number of layers is proven to be universal in the uniform sense for continuous multidimensional input functions as complex coordinates of the quantum state in the computational basis (see \Cref{thm:univ_perez}). We extend this result to fit our needs.

In \Cref{app:oneqbit_univ} we modify universality results from functions as coordinates in the computational basis to functions as the expectation value of an observable. In \Cref{app:uniformtopointwise} we broaden the universality of quantum reuploading models to some discontinuous functions by relaxing the required strength of convergence. More precisely we go from the uniform density in bounded continuous functions to the pointwise density in bounded piece-wise continuous functions. Finally, by stacking $M$ universal circuits, we extend universality to multivariate output functions. All these extensions of \cite{perez-salinas_one_2021} together yield the theorem below.

\begin{theorem}
\label{thm:uniformquniversality}
    For every natural number $M$, for every mapping $f \in \mathcal{B}([0,1]^M \rightarrow [-1,1]^M)$, there exists a sequence of sets of $M$ single qubit quantum circuits and unit norm observables (indexed by $k$).
    \begin{equation}
        \{(U_{k,m} : [0,1]^M \rightarrow \mathcal{U}(2), O_{k,m})_{1 \leq m \leq M}\}_{1\leq k\leq \infty}
    \end{equation}
     such that the sequence of functions $\{g_k\}_{1\leq k \leq \infty}$ defined as 
     \begin{equation}
          g_{k,m}(x) = \bra{0} U_{k,m}(x)^\dag O_{k,m} U_{k,m}(x) \ket{0}
     \end{equation}
     converges pointwise to $f$.
      
      $\mathcal{B}$ is the set of piecewise continuous functions, and the norm of observable is the spectral norm.
\end{theorem}

\begin{figure}
    \centering
    \begin{quantikz}
        \lstick{$\ket{0}$} & \qw & \gate{U_{f_1}(X_1,X_2, \cdots, X_M)} & \qw & \meter{Y_1 \coloneqq \langle Z_1 \rangle}\\
        \lstick{$\ket{0}$} & \qw & \gate{U_{f_2}(X_1,X_2, \cdots, X_M)} & \qw & \meter{Y_2 \coloneqq \langle Z_2 \rangle}\\
        \lstick{$ \vdots $} \\
        \lstick{$\ket{0}$} & \qw & \gate{U_{f_M}(X_1,X_2, \cdots, X_M)} & \qw & \meter{Y_M \coloneqq \langle Z_M \rangle}
    \end{quantikz}
    \caption{Product Encoding Circuit as a universal generator, stacking circuits $U_f$ approximating $f$ from \cite{perez-salinas_one_2021}. It yields the random variable $Y = g(X), X \sim U([0,1]^M)$ with $ f_m=\sqrt{(g_m+1)/2}$.}
    \label{fig.pe_circuit}
\end{figure}
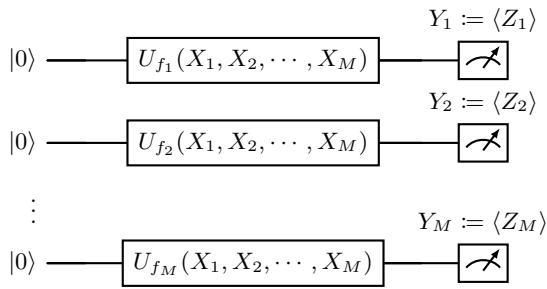

The theorem above shows that there exists a family of $M$-qubit circuits with unit norm observables that yield a family of functions that is pointwise dense in the set of bounded piece-wise continuous functions. This matches the sufficient conditions of \Cref{thm:universalityflow} for mappings to yield a universal generative model. This yields that there exists a family of expectation value sampling models as defined in \Cref{def:EVS} and illustrated in \Cref{fig.pe_circuit} that is universal in the sense of \Cref{def:univgenmod}. 

\begin{theorem}
\label{thm:finitesufficient}
    For any M, for all $M$-dimensional probability density functions $p_Z$ with support included in $[-1,1]^M$, and for all accuracy $\epsilon>0$ there exists a $M$-qubit circuit $U$ and set of $M$ observables $\mathbf{O}$ with unit spectral norm $\norm{O_m}=1$ such that the expectation value sampling model $(U,\mathbf{O},p_X)$ where $p_X$ is the uniform distribution on $[0,1]^M$ yields a probability density function $p_Y$ that is $\epsilon$-close to $p_Z$ in the Wasserstein distance.
\end{theorem}

This is the first central result of this work: $n$-qubit expectation value samplers with constant observable norm are universal for $M$-dimensional distributions with constant support radius, for $M=n$.

\textbf{Note}: We stated our result for distributions on the cube $[-1,1]^M$, but the universality straightforwardly generalizes. In particular, the size of the cube can be rescaled, by rescaling the norm of the observables. Finally, since any distribution with infinite support but finite moments can be arbitrarily approximated by a distribution with finite support, as we show in \Cref{app:infinite_support}, this means that if we allow the observable norm to scale, we can also approximate any distribution with finite moments (but perhaps infinite support).

\subsection{Universality with less qubits}
\label{subsec:univ_de}
 It is worth noting that the construction presented in the previous section uses the same number of qubits as the dimension of the output distribution. However, as we mentioned in the introduction, the dimension of the output does not need to be directly linked to the number of qubits used. This raises the question of the existence of a more qubit-frugal family of universal circuits, in which the output dimension is (much) larger than the qubit number. We will call such constructions "observable-dense expectation value samplers".
 
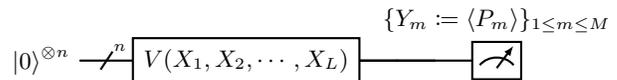
\begin{figure}
    \centering
    \begin{quantikz}
    \lstick{$\ket{0}^{\otimes n}$} & [2mm] \gate{V(X_1,X_2, \cdots, X_L)} \qwbundle{n} & \qw & \qw & \meter{\{Y_m \coloneqq \langle P_m \rangle\}_{1\leq m \leq M}}
    \end{quantikz}
    \caption{Dense Encoding Circuit as a universal generator, based on a universal state preparation circuit $V$, with each parameterized gate replaced by a circuit from \cite{perez-salinas_one_2021}. $n=\log(M+1)$ and $P_m = 2M \ket{m}\bra{m}-I$.}
    \label{fig:de_circuit}
\end{figure}

In \Cref{app:exponentialuniv} we show the existence of such a family if we allow for observables to have large norms. We formalize this in the following theorem and illustrate the corresponding circuit in figure \Cref{fig:de_circuit}.
 
\begin{theorem}
\label{thm:exponentialsufficient}
    For any $M$, for all $M$-dimensional probability density functions $p_Z$ with support included in $[-1,1]^M$, and for all accuracy $\epsilon>0$ there exists a $n=\Theta(\log M)$-qubit circuit $U$ taking $L$ input variables and set of $M$ observables $\mathbf{O}$ with spectral norm $\norm{O_m} \in \Theta(M) $ such that the expectation value sampling model $(U,\mathbf{O},p_X)$ where $p_X$ is the uniform distribution on $[0,1]^L$ yields a probability density $p_Y$ that is $\epsilon$-close to $p_Z$ in the Wasserstein distance.
\end{theorem}

 In this section, we have proven a number of \textit{sufficient} universality conditions for expectation value samplers. In particular, we have shown the existence of two extremal families of parameterized quantum circuits that are universal generators on $[-1,1]^M$. There is a ``product encoding" design, illustrated in \Cref{fig.pe_circuit}, with $n=M$ qubits and with unit norm observables (local Pauli), and a ``dense encoding" design, illustrated in \Cref{fig:de_circuit} with $n=\log(M)$ qubits and $M$ norm observables (amplified probabilities of bitstrings). 
 The product encoding circuit has a large number of qubits for a constant observable norm, and the dense encoding circuit has a logarithmic number of qubits but large observable norms. This hints that there might be a trade-off between the number of measurements and the number of qubits. In the next section, we prove this is the case, by proving some \textit{necessary} universality conditions.

\section{Necessary conditions for universality}
\label{sec:necessary}
In this section, we use Holevo-like bounds to prove some necessary conditions on the number of qubits and the observable norm for an expectation value sampling model to be universal on $[-1,1]^M$. 

\subsection{Dimension of the observables space}
\label{subsec:nec_obsdim}
As previously mentioned, an appealing feature of expectation value sampling models is that the dimension of the output vector is \textit{a priori} independent of the number of qubits $n$, unlike in the case of quantum Born machines where each qubit corresponds to exactly one binary random variable. In particular, we can imagine using even just a single qubit with an arbitrary number of observables ${O_m}$ to generate an $M$-dimensional random vector. However, it is obvious that in this case, the random variables corresponding to each observable cannot all be fully independent. Indeed, any observable can be expressed as a linear combination of the three Pauli matrices. Therefore the distribution output by a single qubit expectation value sampler will have at most three degrees of independence, and for $M>3$ it is impossible to have universality because it is impossible to approximate e.g. the 4-dimensional uniform distribution. Extending this reasoning to several qubits, we find the first necessary condition, that the dimension of the target dimension has to be lower or equal to the dimension of the space of observables. We formalize this in the following theorem.

\begin{theorem}
     For an $n$-qubit expectation value sampling model $(U_{\theta}(x),\mathbf{O},p_X)$ to be able to approximate any distribution with support in $[-1,1]^M$ to any accuracy $\epsilon>0$, it is necessary that $M\leq 4^n-1$.
\end{theorem}

\subsection{Holevo's bound}
\label{subsec:nec_holevo}
Another necessary condition for an expectation value sampling model to be universal on distributions with support in $[-1,1]^M$ can be derived using a combination of Holevo's bound found in \cite{ambainis_dense_2002} and Chernoff bound. We formalize it in the following theorem, which we prove in \Cref{app:proof_nec}.

\begin{theorem}
    For an $n$-qubit expectation value sampling model $(U_{\theta},\mathbf{O},p_X)$ to be able to approximate any distribution with support in $[-1,1]^M$ to any accuracy $\epsilon>0$ with respect to the Wasserstein distance, it is necessary that for every $m \leq M$:
    \begin{enumerate}
        \item $ \lm(O_m) \leq -1+\epsilon$ and $\lM(O_m) \geq +1-\epsilon$
        \item $ n \in \Omega\left(\frac{M (1-\epsilon)^2}{\norm{O_m}^2}\right)$
    \end{enumerate}
    with $\lambda_{\min / \max}(O)$ returning respectively the minimum and maximum eigenvalues of observable $O$.
    \label{thm:necessary}
\end{theorem}
The above necessary conditions use results which involve a more general case where we assume that prior to measurement we use a general parameterized quantum channel which can also prepare mixed states. However, we can reduce this to the special case of unitaries as well. By purification, mixed states can be mimicked by using $2n$ qubits pure states, since we are only interested in scalings, the factor $2$ plays no role in the second condition of \Cref{thm:necessary}, and the necessary conditions also hold for pure states.

The combination of both previous necessary conditions is the second central result of this paper. It formalizes that even if expectation value sampling models may output arbitrary large dimensional distributions, in practice their expressivity is limited by the number of qubits and observables. 

\subsection{Asymptotic optimality of the product and dense encoding families}
\begin{figure}
    \centering
    \includegraphics[width=\columnwidth]{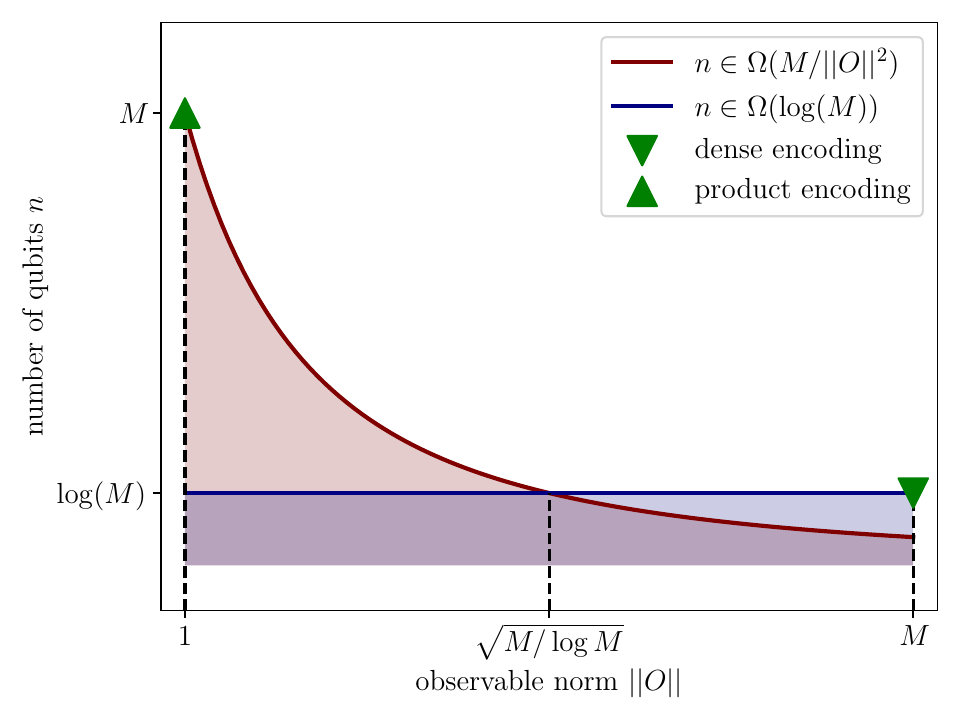}
    \caption{Visual summary of results. We show the asymptotic use of resources for expectation value samplers to reach universality for a $M$-dimensional target distribution: the necessary conditions from \Cref{subsec:nec_obsdim} and \Cref{subsec:nec_holevo}, as well as the existence of families from \Cref{subsec:univ_pe} and \Cref{subsec:univ_de}.}
    \label{fig:visusum}
\end{figure}

It is a natural question to ask whether the universal families we found in \Cref{sec:sufficient} saturate the necessary conditions found in the previous subsection. We illustrate such considerations in \Cref{fig:visusum}.

\begin{corollary}
    The product encoding family defined in \Cref{subsec:univ_pe} has a number of qubits $n \in \Theta(M)$ and an observable norm of $1$. It is optimal with respect to the necessary condition from \Cref{subsec:nec_holevo}.
\end{corollary}

\begin{corollary}
    The dense encoding family defined in \Cref{subsec:univ_de} has a number of qubits $n \in \Theta(\log(M))$ and an observable norm of $\norm{O} \in \Theta(M)$. It asymptotically saturates the condition from \Cref{subsec:nec_obsdim}. However, it is not optimal with respect to the condition from \Cref{subsec:nec_holevo}, that yields $\norm{O} \in \Omega(\sqrt{M/\log(M)})$, leaving a quasi-quadratic gap. 
\end{corollary}

 We may conjecture that there exists a family of (Pareto) optimal circuits, balancing between observable norms and qubit numbers, with varying ``encoding densities" in between the two extremal cases. In practice, the observable spectral norm relates to the number of measurements required to approximate it up to an additive accuracy. This highlights a trade-off between the number of qubits and the number of measurements, or space versus time complexity, which we formalize in the next subsection. 

\subsection{Observable norm and number of measurements}
To estimate an expectation value up to a desired constant additive error, the number of measurements required is proportional to the norm of the observables. Therefore, large observable norms require a large number of measurements. This intuition is sharpened in the following lemma, proven in \Cref{app:approxexp}.
\begin{lemma}
    An arbitrary expectation value sampler outputs an $M$-dimensional random vector $Y \sim p_Y$ with an infinite number of measurements, i.e. with access to exact expectation values. We consider the same circuit but with a finite number of measurements $T$ that estimates expectation value by sampling and averaging for each observable and yields a random vector $\hat{Y} \sim p_{\hat{Y}}$. The number of measurements $T$ required to guarantee that the Wasserstein distance between both distributions is smaller than $\epsilon$ satisfies
    \begin{equation}
        T \in \Theta\left(\frac{M \norm{O}}{\epsilon^2}\right).
    \end{equation}
\end{lemma}

This lemma can be applied to both the product encoding circuit and the dense encoding circuit to show the trade-off between the number of qubits and the number of measurements.

\subsection{On correlations in the target distribution}
It is interesting to note that the parameter driving the necessary conditions is not the dimension of the target distribution itself $M$ but rather the number of independent variables $M'$. Because we defined universality as being able to capture full independence we use $M=M'$ in the proofs of the necessary conditions. However, by restricting our interest to universality for classes of distributions with certain dependence conditions fulfilled, we can obtain much more economic bounds than for the general case, which is illustrated in \Cref{fig:visusum}. We give two examples of universality for a restricted family of distributions.

First, consider the family of distributions that are $M$-dimensional but that have their support included in a $2$-dimensional linear subspace. We choose the expectation value sampler as the circuit from \cite{perez-salinas_one_2021} and observables $\{\sqrt{2} X, \sqrt{2} Z\}$ we have a universal generator on one qubit for this $2$-linear family for \textit{any} target distribution dimension $M$.

Second, consider the family of $M$-dimensional Dirichlet distributions, for which each coordinate is positive, and such that the sum of all coordinates is unity. For this case, we can take the ``dense encoding" circuit with $n = \log{M}$ qubits, see \Cref{fig:de_circuit}, and instead of amplified probabilities we take the raw probabilities as observables. This constitutes a universal model for the Dirichlet family. In this case, we get away with exponentially fewer qubits than the general case. 

These observations may be used to suggest that expectation value samplers are better suited to generate highly correlated distributions. Note that for the Dirichlet case, the result is not surprising as the normalization condition perfectly matches the normalization of quantum states, making this type of universality particularly well suited to expectation value samplers.

\section{Characterizing expectation value sampling distributions}
In this section, we propose additional tools to refine our understanding of expectation value sampling models, exploring how the choice in encoding, input distribution, and observables affects the expressivity.

\subsection{Observable choice and primary mapping} 
In this subsection we explore the expressivity that comes with the choice in observables, considering expectation value models with fixed encoding $U(x)$ on $n$ qubits and $M$-dimensional inputs. We are using the standard Pauli basis for the space of $2^n \times 2^n$ Hermitian operators $\mathbf{P}$. It is composed of all possible combinations of $n$ Pauli matrices $\sigma_{0,1,2,3}$, which yields $|\mathbf{P}| = 4^n$. We formalize as follows
\begin{align}
    \mathbf{P}  \coloneqq & (P_k, k \in \{0,1,2,3\}^n) \\
    = & (\otimes_{1 \leq i \leq n} \sigma_{k_i}, k_i \in \{0,1,2,3\}).
\end{align}

Any vector of $M$ observables $\mathbf{O} = (O_m)$, can be expressed as a linear mapping applied on the vector of all Pauli strings: $\mathbf{O} = A \mathbf{P}$, where $A$ is an $M \times 4^n$ matrix, and $\mathbf{P}$ is a $4^n$ dimensional vector. Therefore the distribution associated with the Pauli basis encompasses any distributions, which leads us to define the primary mapping as follows.

\begin{definition}[Primary mapping]
\label{def:primarymapping}
    The primary mapping $g$ of an $n$-qubit encoding circuit $U_{\theta}(x)$ is defined as the mapping of the associated expectation value sampling model with the Pauli basis $\mathbf{P}$, defined as $(U_{\theta}(x),\mathbf{P},p_X)$ according to \Cref{def:EVS}. It can be expressed as follows
    \begin{equation}
        x \in [0,2\pi)^N \xrightarrow{g} \left( \bra{0} U_{\theta}(x)^\dag P_k U_{\theta}(x) \ket{0} \right)_{1 \leq k \leq 2^n}.
    \end{equation}
    It yields the $4^n$-dimensional random variable $Z = g(X)$ when $X \sim p_X$.
\end{definition}

It is easy to see that any distribution obtained by an expectation value sampling model can always be expressed by considering an intermediary output of the $4^n$ set of observables, followed by the linear mapping $A$. We capture this idea in the following theorem.

\begin{theorem}
\label{thm:linearobservables}
    Given an encoding circuit $U_{\theta}(x)$ and random variable with distribution $p_X$, for any the choice of observables $\mathbf{O}$, the expectation value sampling model $(U_{\theta}(x),\mathbf{O},p_X)$ (according to \Cref{def:EVS}) is a linear transformation of the expectation value sampling model $(U_{\theta}(x),\mathbf{P},p_X)$, where $\mathbf{P}$ is the Pauli basis per definition of the primary mapping in \Cref{def:primarymapping}.
\end{theorem}

This concept of primary mapping has immediate consequences on the possible correlation of output variables of expectation value sampling models. For a given data encoding part on $n$ qubits, the primary mapping will yield a random variable $Z$ with a covariance matrix $C_z$. Because any expectation value sampling model based on the same data encoding is a linear transformation of the primary mapping, the number of uncorrelated variables is limited by the number of non-null eigenvalues of the covariance matrix $C_z$, which is upper bounded in any case by $4^n-1$.
\begin{lemma}
    Given an encoding circuit $U(x)$ over $n$ qubits and random variable with distribution $p_X$, we note $L$ the number of non-zero eigenvalues of the covariance matrix of the primary mapping. For any choice of observables, any expectation value model using $(U(x),p_X)$ will yield at most $L$ uncorrelated variables. In addition $L \leq 4^n-1$.
\end{lemma}

\subsection{Random variable encoding as a polynomial chaos expansion}

After focusing on the choice of observables, in this subsection, we analyze the impact of the choice of the input random variable and the circuit encoding on the expressivity. We propose the polynomial chaos expansion \cite{xiu_numerical_2010} as a useful tool to analyze the expressivity of expectation value sampling models, as the analogue of the Fourier decomposition. The general polynomial chaos expansion is a representation of random variables as a vector in a Hilbert space of orthogonal functions, as defined below.

\begin{definition}[Generalized Polynomial Chaos Expansion]
    A generalized chaos expansion is characterized by a probability density function $p_X$ defined on the support $\mathcal{X} \subseteq \mathbb{R}^M$ with finite moments (usually chosen as standard distributions, such as Gaussian or uniform). This choice defines an inner product for functions in $\{f : \mathcal{X} \rightarrow \mathbb{R}\}$:
    \begin{equation}
        \label{eq:innerproduct}
        {\bra{f}\ket{g}}_{p_X} \coloneqq \int_{\mathcal{X}} f^*(x) g(x) p_X(x) dx.
    \end{equation}
    This choice of inner product comes with the choice of an ordered family of functions, usually polynomials, that are orthonormal with respect to the above inner product. 
    \begin{equation}
        \Phi_{p_X} = \{\phi_l : \mathcal{X}\rightarrow\mathbb{R}, \forall(k,l), \ket{\phi_l} \bra{\phi_k} = \delta_{k,l}\}
    \end{equation}
    A generalized chaos expansion is a representation of a random variable $Y$ with probability density $p_Y$ as a vector $\alpha$ in this Hilbert space, such that:
    \begin{equation}
    \label{eq:pce}
        Y = \sum_{l=0}^{\infty} \alpha_l \phi_l(X) \sim p_Y, X \sim p_X
    \end{equation}
\end{definition}
This provides a Hilbert space as a potential structure to study random variable mappings. In particular, one of the main results of polynomial chaos expansion is that they are universal generative models. 

Common pairs of distribution and associated orthogonal polynomials family can be found in \Cref{tab:poly_distrib}. In the context of expectation value sampling, we focus on the family of functions  orthonormal with respect to the inner product associated with the uniform distribution on $\mathcal{X} = [0,2 \pi)^M$
\begin{equation}
    \Phi = \{\prod_{1 \leq m \leq M} e^{i k_m x_m}, k \in \mathbb{Z}^M\}
\end{equation}

Quantum reuploading circuits are a widely used class of parameterized quantum circuits that output a function of the data. They are used in a regressive context, where optimization techniques are used such that their output fits a target function. It has been widely studied and used that if they use integer-valued spectrum Hamiltonian, their output hypothesis function can be decomposed as an exact finite Fourier series \cite{schuld_effect_2021}. This means that there exists $c_{\mathbf{k},l} \in \mathbb{C}$ such that the hypothesis function $f$ can be exactly written as a finite Fourier series:
\begin{equation}
    f(x) = \sum_{k_0,k_1,\cdots,k_M = -K}^{+K} c_{\mathbf{k},l} \prod_{1 \leq m \leq M} e^{i k_m x_m}, .
\end{equation}

This fact extends to a generative modelling context where expectation value sampling models using integer-valued quantum reuploading circuits yield distributions with an exact finite polynomial chaos expansion. We formalize this below.
\begin{theorem}
    \label{thm:fourierpce}
    Any expectation value sampling model using a quantum reuploading model with integer-valued spectrum $U_{\theta}(x)$, together with the uniform distribution on $[0,2\pi)^M$ outputs a random variable $Y$ that has an exact finite polynomial chaos expansion for any choice of observables.
\end{theorem}
This subsection formalizes the tight connection between quantum circuits used in a regressive context with their use in a generative context. Therefore it is expected that, beyond the universality, many properties of such models can be transferred from a regressive context to a generative context, but we leave that for future work.
\begin{table}[]
    \centering
    \begin{tabular}{ |c|c| } 
        \hline
        Normal distribution & Hermite polynomials \\ 
        \hline
        Uniform distribution & Legendre polynomials \\ 
        \hline
        Exponential distribution & Laguerre polynomials \\ 
        \hline
        Beta distribution & Jacobi polynomials \\ 
        \hline
    \end{tabular}
    
    \caption{Pairs of distributions and corresponding orthonormal families commonly used in General Polynomial Chaos expansion.}
    \label{tab:poly_distrib}
\end{table}

\subsection{Choice of input random variable}
We discuss the choice of input random variable, noted as $p_X$ in \Cref{def:EVS}. We have shown several reasons to consider the uniform distribution as a good design choice as an input random variable, mostly based on the fact that it has a bounded support. Indeed, it is known that the most commonly used parameterized quantum circuits are periodic, in particular when they have a finite Fourier decomposition. This is also why the universality of quantum reuploading circuits is proven for functions on bounded domains, corresponding to a half period. In contrast, let us consider an expectation value sampler $(U,\textbf{O},p_X)$, where the input random variable follows a Gaussian distribution $X \sim \mathcal{N}(0,1)$, which has unbounded support and for which the mapping $f$ is $1$-periodic. $X=0$ is the highest probability event and will yield the exact same output as a very low probability event, for example, $X=100$. This means that a very low probability input event and a very high probability input event will yield the same output sample, which is a feature rarely considered desirable. This choice of uniform distribution is in contrast to classical GANs where Gaussians are typically preferred.

\vspace{20px}

\section{Contributions and discussion}
The first central result of this work is that expectation value samplers, based on parameterized quantum circuits, are universal generative models. We provide constructive proofs for two extremal circuit designs. The ``product encoding" design has a number of qubits linear in $M$ and unit norm observables. In contrast, the ``dense encoding" design has a number of qubits logarithmic in $M$ and observable norm linear in $M$. The second central result is necessary conditions on resources such as the number of qubits and number of measurements for expectation value sampling models to be universal. This allows us to assess the optimality of the universal family we constructed. 
We conjecture there exists a series of universal circuits with varying ``encoding densities" in between these two extremal cases, that are optimal with respect to the necessary conditions we have proven in this work. In addition, we propose additional tools to analyse the expressivity of expectation value samplers. We hope that the knowledge presented in this work will guide the choice of random variable encoding and observables in future designs.

While we characterized important properties of expectation value samplers, we did not address the question of whether it is a good idea to use them. Expectation value samplers cannot be proven to be a path for certain types of quantum advantage as easily as is the case in Quantum Circuit Born Machines \cite{coyle_born_2020}. Indeed, expectation value samplers, rather than requiring sampling from the full distribution of quantum measurements, are defined around expectation values only. Thus the hardness of simulating distributions from expectation value samplers does not connect straightforwardly to any hardness-of-sampling results established in the domain of quantum supremacy results \cite{coyle_born_2020}.

Nonetheless, expectation value samplers still consist of genuinely quantum computations, and arguments for non-simulatability can be made. Assuming BQP is not in BPP, there exists no polynomial time algorithm that takes a classical description of an arbitrary expectation value sampler $A$ as an input and outputs a sample from a distribution that is epsilon close to that of the output of $A$. Take as an example the case where a hard-to-simulate circuit does not depend on input data. This yields a Dirac delta distribution for which there exist classical samplers to efficiently sample from it, however, such classical samplers cannot be easily found based on the classical description of the expectation value sampler. It may be possible to construct stronger advantage arguments where we find the existence of an expectation value sampler such that its output distribution cannot be sampled by any polynomial-time randomized Turing machine (subject to standard assumptions) but we leave this for future work.

\acknowledgements
AB thanks Adrián Pérez-Salinas for helpful clarifications and discussions on the universality of quantum circuits. AB and MG are supported by CERN through the CERN Quantum Technology Initiative. AB is supported by the quantum computing for earth observation (QC4EO) initiative of ESA $\phi$-lab, partially funded under contract 4000135723/21/I-DT-lr, in the FutureEO program. This work was supported by the Dutch National Growth Fund (NGF), as part of the Quantum Delta NL programme. This work was supported by the Dutch Research Council (NWO/ OCW), as part of the Quantum Software Consortium programme (project number 024.003.037). This work was supported by the European Union’s Horizon Europe program through the ERC StG FINE-TEA-SQUAD (Grant No. 101040729) and through the ERC CoG BeMAIQuantum (Grant No. 101124342). JT has received support from the European Union’s Horizon Europe research and innovation programme.
Views and opinions expressed are however those of the author(s) only and do not necessarily reflect those of the European Union, the European Commission, or the European Space Agency, and neither can they be held responsible for them.

\bibliographystyle{alpha}
\bibliography{references}

\vspace{12pt}

\appendix

\section{Constructive mapping between a random variable and the uniform distribution}
\label{app:triangular}
    Let us consider an absolutely continuous random variable $Y$ with probability density function $p_Y$ on a bounded set $[a,b]^M$.
    We recall a definition of an absolutely continuous variable below.
    \begin{definition}
    \label{def:acrv}
        A random variable $X$ is said to be absolutely continuous if its cumulative distribution function (CDF) can be expressed as the integral of a non-negative function, known as the probability density function (PDF).
    \end{definition}
    This excludes for example Dirac deltas. We are going to construct an invertible mapping to transform the uniform random variable $X$ into this random variable $Y = [Y_k]$. We call $G_1 : [0,1] \rightarrow [a,b]$ the cumulative distribution function of the marginal of $Y_1$. It is invertible and we define $F_1$ as its inverse,
    \begin{equation}
    Y_1 = F_1(X_1), X_1  = G_1(Y_1).
    \end{equation}
    Next we consider the marginal of $Y_2$ conditioned by $Y_1$, we define the cumulative distribution $G_2 : [a,b]^2 \rightarrow [0,1]$,
    \begin{equation}
     G_2(y_1,y_2)=P(Y_2=y_2|Y_1=y_1).
    \end{equation}
    It is invertible with respect to $Y_2$,
    \begin{equation}
     G_2^{-1}(Y_1,X_2)=Y_2 \iff G_2(Y_1,Y_2)=X_2.
    \end{equation}
    We define $ F_2: [0,1]^2 \rightarrow [a,b]$ as follows,
    \begin{equation}
    F_2(X_1,X_2) = G_2^{-1}(F_1(X_1),X_2).
    \end{equation}
    We have $Y_2 = F_2(X_1,X_2)$. Continuing this process iteratively for all coordinates, it is possible to fully define the invertible mapping $F = [F_k]$ with inverse $G=[G_k]$ such that $Y=F(X)$. This is the essence of triangular mapping in \cite{bogachev_triangular_2005}. In addition, this mapping is bounded and piece-wise continuous, because it is composed of inverse cumulative distribution functions of absolutely continuous variables on bounded support. 

\section{From state coordinate universality to expectation of observable universality}

\label{app:oneqbit_univ}
We start by recalling Theorem 4 from \cite{perez-salinas_one_2021}, which proves that the following circuit is universal. It has $L$ layers, $\theta \in \mathbb{R}^{(M+2)\times L}$ parameters and for $x\in\mathbb{R}^M$ is defined as
\begin{equation}
    U_{\theta}(x) \coloneqq \prod_{l=1}^L R_y(\theta_{0,l}) \left( \prod_{m=1}^M R_z(x_m \theta_{m,l}) \right) R_z(\theta_{M+1,l}).
\end{equation}

\begin{theorem}
\label{thm:univ_perez}
    For any pair of functions and real number
    \[(f \in \mathcal{C}([0,1]^M\rightarrow[0,1]) , \phi \in \mathcal{C}([0,1]^M\rightarrow[0,2\pi)) ,\epsilon>0)\]
    There exists a one qubit circuit $U:[0,1]^M \rightarrow \mathcal{U}(2)$ s.t.
    \begin{equation}
        \forall x, \abs{\bra{1} U(x)\ket{0} - f(x) e^{i \phi(x)}} < \epsilon.
    \end{equation}
\end{theorem}
In this work, $\mathcal{C}$ is the set of continuous functions.
This theorem yields the universality of functions embedded in a quantum state in the uniform sense. In the context of expectation value sampling, we are interested in the universality of function as the expectation value of a unit norm observable, captured by the following theorem.

\begin{theorem}
    For any function $g \in \mathcal{C}([0,1]^M\rightarrow[-1,1])$ and for any $\epsilon>0$, there exists a one qubit circuit $U(x):[0,1]^M \rightarrow \mathcal{U}(2)$ and an observable $O$ with unit spectral norm $\norm{O}=1$ s.t.
    \begin{equation}
        \forall x, \abs{ \bra{0} U^{\dagger}(x) O U(x)\ket{0} - g(x)} < \epsilon.
    \end{equation}
    \label{thm:univunifexpect}
\end{theorem}

\textit{Proof:} We are given an arbitrary function $g \in \mathcal{C}([0,1]^M\rightarrow[-1,1])$ and $\epsilon>0$. We define the function $f=\sqrt{\frac{g+1}{2}}$ which is well defined on $\mathcal{C}([0,1]^M\rightarrow[0,1])$. We apply \Cref{thm:univ_perez} to $(f,\phi=0,\epsilon/4)$ and get a circuit $U$ that yields a state close to 
\begin{equation}
    \ket{x} = \sqrt{1-f(x)^2} \ket{0} + f(x) \ket{1}.
\end{equation}
The $Z$ expectation value of the above state is 
\begin{equation}
    \bra{x} Z \ket{x} = 2 f(x)^2 - 1 = g(x).
\end{equation}
We are now going to prove that the expectation value is close to the target function $g$. First, we note that the square function is $2$-Lipschitz on $[0,1]$ and therefore for every pair of real numbers $(x,y)\in [0,1]^2$
\begin{equation}
    \abs{x-y} < \epsilon \implies \abs{x^2-y^2} < 2 \epsilon.
\end{equation}
We define $p_{0/1}$ the probabilities of measuring $U(x) \ket{0}$ in state $\ket{0}$ and $\ket{1}$ respectively. Recalling that 
\begin{equation}
    \abs{\bra{1} U(x) \ket{0} - f(x)} < \epsilon/4,
\end{equation}
we can write
\begin{align}
    \abs{p_1 - f(x)^2} &< \epsilon/2\\
    \abs{p_0-(1-f(x)^2)} &=\nonumber \\ 
    \abs{1-\abs{\bra{1} U(x) \ket{0}}^2 - (1-f(x)^2)} &< \epsilon/2.
\end{align}
Finally,
\begin{equation}
    \abs{\langle Z \rangle - g(x)} \leq \abs{p_1 - f(x)^2} + \abs{p_0-(1-f(x)^2)} < \epsilon.
\end{equation}
This yields the uniform density of quantum functions in the set of bounded continuous functions. 

\section{From uniform density on continuous functions to pointwise density on discontinuous functions}
\label{app:uniformtopointwise}
We first start by highlighting the difference between uniform convergence and pointwise convergence and illustrate it with an example.

\begin{definition}[Pointwise convergence]
    Let $\{f_k | f_k : \mathcal{X} \rightarrow \mathbb{R}\}_{1\leq k \leq \infty}$ be a sequence of functions, and let $f : \mathcal{X} \rightarrow \mathbb{R}$ be another function defined on the same domain $\mathcal{X}$. We say that the sequence $\{f_k\}$ converges pointwise to $f$ if, for each $x \in \mathcal{X}$, the sequence of real numbers $\{ f_k(x) \}_{1\leq k \leq \infty}$ converges to $f(x)$ as $k$ approaches infinity,
    \[
    \lim_{n \to \infty} f_k(x) = f(x) \quad \text{for all } x \in \mathcal{X}.
    \]
\end{definition}

\begin{definition}[Uniform convergence]
    Let $\{f_k | f_k : \mathcal{X} \rightarrow \mathbb{R}\}_{1\leq k \leq \infty}$ be a sequence of functions, and let $f : \mathcal{X} \rightarrow \mathbb{R}$ be another function defined on the same domain $\mathcal{X}$. We say that the sequence $f_k$ converges uniformly to $f$ if, for any given $\epsilon > 0$, there exists an $K \in \mathbb{N}$ such that for all $k \geq K$ and for all $x \in \mathcal{X}$, the difference $|f_k(x) - f(x)|$ is less than $\epsilon$,
    \[
    \forall \epsilon > 0, \exists K \in \mathbb{N} : \forall k \geq K, \forall x \in \mathcal{X}, |f_k(x) - f(x)| < \epsilon.
    \]
\end{definition}

Uniform convergence is stronger, it implies pointwise convergence, but the reverse is not true. For example, consider the step function $f$. 
\begin{equation}
    f(x)=
    \begin{cases}
      -1, & \text{if} \ x \in [-1,0[ \\
      +1, & \text{if} \ x \in [0,+1]
    \end{cases}
\end{equation}

\begin{figure}
    \centering
    \includegraphics[width=\linewidth]{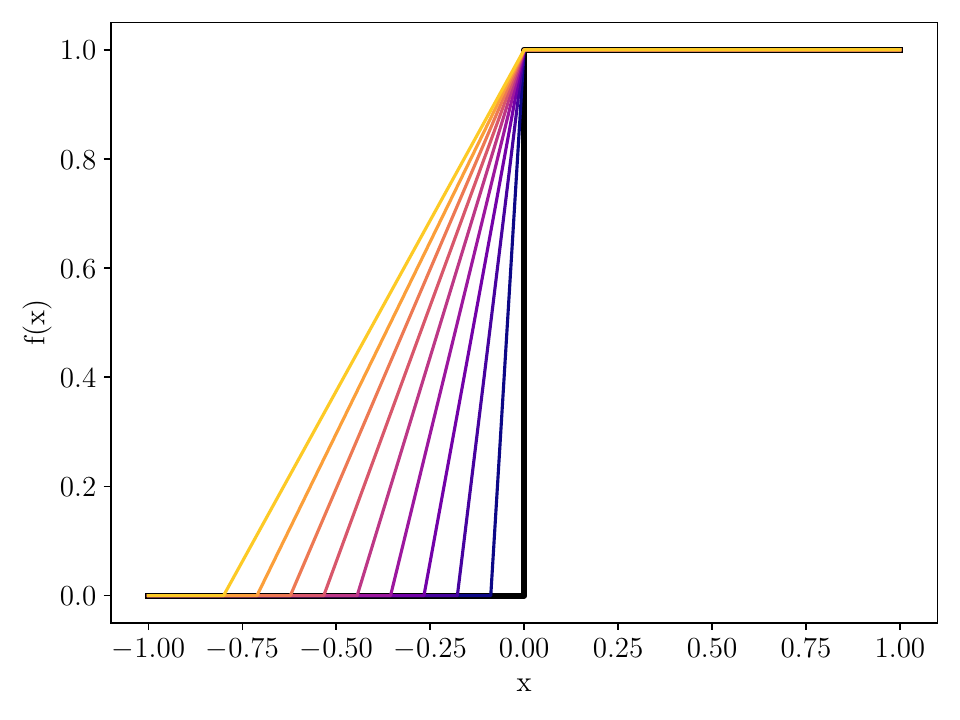}
    \caption{A sequence of continuous functions converging pointwise but not uniformly to the step function.}
    \label{fig:pwstep}
\end{figure}

It is impossible to define a sequence of continuous functions that would uniformly converge to it, however, it is possible to have a sequence of continuous functions that converges pointwise to it, see \Cref{fig:pwstep}.
\begin{equation}
    f_k(x)=
    \begin{cases}
      -1, & \text{if} \ x \in [-1,1/k[ \\
      1+kx, & \text{if} \ x \in [-1/k,0] \\
      +1, & \text{if} \ x \in ]0,+1]
    \end{cases}
\end{equation}

In \cite{baire_sur_1899} Baire defined hierarchical pointwise convergence classes of functions. Baire class 0 is the set of continuous functions, and the class $c$ is the set of functions that are the pointwise limit of class $c-1$. In particular in \cite{laczkovich_baire_1984, lebesgue_propriete_1904}, it was proven that bounded piece-wise continuous functions are of class $1$. This means that for any bounded piece-wise continuous function $f$ there exists a sequence of bounded continuous functions that converges pointwise to $f$. This means that bounded continuous functions are dense in bounded piece-wise continuous functions in the pointwise topology. Therefore, building on \Cref{thm:univunifexpect} we have the following theorem, writing $\mathcal{B}$ as the set of piecewise continuous functions.
\begin{theorem}
    For any function $f:\mathcal{B}([0,1]^M\rightarrow[-1,1])$, there exists a sequence (indexed by $k$) of one qubit circuit and observables with unit spectral norm,
    \begin{equation}
        \{(U_k(x):[0,1]^M \rightarrow \mathcal{U}(2),O_k)\}_{1\leq k\leq\infty}
    \end{equation}
    such that the sequence of functions $\{g_k\}$ with
    \begin{equation}
        g_k(x) = \bra{0} U_k^{\dagger}(x) O_k U_k(x)\ket{0}
    \end{equation}
    converges pointwise to $f$. 
\end{theorem}

\section{Notes on infinite support}
\label{app:infinite_support}
We consider a distribution with probability density function $p_Y$ with infinite support $\mathbb{R}^M$ and finite moments. The goal is to find a sequence of random variables with bounded support probability density function that converges pointwise to $p_y$. We define the sequence of random variables $\{Y_k\}_{1\leq k \leq \infty}$ whose probability density functions $p_{Y_k}$ are proportional to that of $Y$ on the hypercube $[-k-k_0,+k+k_0]^M$, where $k_0$ is the first integer such that $Y$ has non-null support on the hypercube. This sequence converges in distribution to $Y$.

\section{Construction of the ``dense encoding" circuit}
\label{app:exponentialuniv}
\textbf{Note}: We have used slight abuse of notations in the below proof to increase readability, specifically in the approximations.

Considering architecture such as in \cite{plesch_quantum-state_2011} and universal one qubit gate as $R_x(\alpha) R_z(\beta) R_x(\gamma)$, for any number of qubits, it is possible to design a circuit $U: [0,2\pi)^L \rightarrow \mathcal{U}(2^n)$ made only of a finite number of fixed gates (CNOT and constant rotations) and parameterized $\sigma_z$ rotations gates that can reach any pure quantum state when applied to state $\ket{0}$,
\begin{equation}
    \forall \ket{\psi}, \exists \theta \in [0,2\pi)^L, \ket{\psi} = U(\theta) \ket{0}.
\end{equation}
Let's consider a distribution $p_{\psi}$ over pure states. Because the architecture above can reach any pure state, there exists a corresponding distribution over the parameters $p_\theta$ such that the distribution $V(\theta) \ket{0}, \theta \sim p_{\theta}$ matches perfectly $p_{\psi}$ in distribution. We define $g:[0,1]^L\rightarrow[0,2\pi)^L$ as the mapping that transforms the uniform distribution over $[0,1]^L$ into $p_\theta$. We have $V(f(X)) \ket{0}, X \sim U([0,1]^L)$ matches perfectly $p_{\psi}$ in distribution, where $V$ is composed of a finite number of fixed gates and $L$ $\sigma_z$ rotation gates parameterized by $g_l(X)$. 

\begin{lemma}
    For any distribution over pure states $p_{\psi}$, there exists a circuit architecture $V$ made of constant gates and parameterized $\sigma_z$ rotations, and a mapping $g$ such that 
    \begin{equation}
        X \sim U([0,1]^L), V(g(X)) \ket{0} \sim p_{\psi}
    \end{equation}
\end{lemma}

Next, we decompose $R_z \circ g$ gates into a sequence of constant gates and parameterized $\sigma_z$ rotations.

From \cite{perez-salinas_one_2021}, in the proof in the appendix, it is shown that there exists a quantum circuit $W$ taking a multidimensional input and approximating the following parameterized quantum gate,
\begin{multline}
    \forall f:[0,1]^L\rightarrow [0,1], \phi:[0,1]^L\rightarrow [0,2 \pi), \exists W\\
    W(x) \equiv_{\epsilon} 
    \begin{bmatrix}
        \sqrt{1-f(x)^2} e^{+i\phi(x)} & -f(x) e^{+i\phi(x)} \\
        f(x) e^{-i\phi(x)} & \sqrt{1-f(x)^2} e^{-i\phi(x)} 
    \end{bmatrix}.
\end{multline}
Choosing $\phi=0$, and $f=\sin{g}$, we have $\sqrt{1-f^2}=\cos{g}$, and we note that $W(x)=R_z \circ g$. We can conclude the following.
\begin{lemma}
    \begin{equation}
        \forall f:[0,1]^L\rightarrow [0,1], \exists W, R_z \circ f \equiv_{\epsilon} W,
    \end{equation}
    where $W$ is a quantum circuit made of constant gates and $R_z$ gates applied to individual components of $x$.
\end{lemma}

Combining both above lemmas we get the following.
\begin{lemma}
     For any distribution over pure states $p_{\psi}$, there exists a circuit architecture $V$ made of constant gates and parameterized $z$ rotations such that 
    \begin{equation}
        X \sim U([0,1]^L), V(X) \ket{0} \sim_{\epsilon} p_{\psi}.
    \end{equation}
\end{lemma}

We are now going to use that lemma to prove that $n$-qubits expectation value samplers are universal for $\exp(n)$-dimensional distributions with constant support if the observables are allowed to have $\exp(n)$ norms. 

We are given an arbitrary random variable $Y$ following a $M$-dimensional distribution $p_Y$ with support $[-1,1]^M$. We define the following state over $n$ qubits with $M=2^n-1$.
\begin{align}
    \ket{\psi(Y)} &= \sum_{m\leq M} Z_m \ket{m} + \sqrt{1-\sum_{m\leq M} Z_m^2} \ket{M+1}\\
    Z_m &=\sqrt{\frac{Y_m+1}{2 M}}
\end{align}

We define as $p_{\psi}$ as the probability density functions over states when $Y\sim p_Y$, using the previous lemma we get a circuit $W$ composed only of constant gates and $z$ rotations gates with one of the $L$ parameters as input that approximates $p_{\psi}$. We define the observables $\forall m\leq M, O_m = 2 M \ket{m} \bra{m} - I$. They have spectral norm $\norm{O_m} = 2M-1 = \Theta(2^n)$. In addition, $\bra{\psi(Y)} O_m \ket{\psi(Y)} = Y_m$. Therefore, the expectation value sampler $(W,O,U([0,1]^L))$ approximates $p_Y$. This concludes the proof to \Cref{thm:exponentialsufficient}.

\section{Proof of theorem in \Cref{subsec:nec_holevo}}
\label{app:proof_nec}
We start by proving the \Cref{thm:necessary}, which we recall below.
\begin{theorem*}
    For an $n$-qubit expectation value sampling model $(U_{\theta},\mathbf{O},p_X)$ to be able to approximate any distribution with support in $[-1,1]^M$ to any accuracy $\epsilon>0$ with respect to the Wasserstein distance, it is necessary that for every $m \leq M$:
    \begin{enumerate}
        \item $ \lm(O_m) \leq -1+\epsilon$ and $\lM(O_m) \geq +1-\epsilon$
        \item $ n \in \Omega\left(\frac{M (1-\epsilon)^2}{\Lambda(O_m)}\right) \subseteq  \Omega\left(\frac{M (1-\epsilon)^2}{\norm{O_m}^2}\right)$
    \end{enumerate}
    with $\lambda_{\min / \max}(O)$ returning respectively the minimum and maximum eigenvalues of observable $O$, and $\Lambda(O)\coloneqq -\lambda_{\min}(O)\lambda_{\max}(O)$.
\end{theorem*}

\textit{Proof:} Let's suppose there is an $n$-qubit expectation value scheme with $M$ observables $\mathbf{O}$ that is able to approximate any distributions with support included in $[-1,1]^M$ to $\epsilon$ with respect to the first Wasserstein distance $W_1$.

Because the expectation value model is universal, it means that for any vertex of the hypercube $c \in \{-1,1\}^M$, it can approximate the Dirac delta at $c$. We then use the lemma below.
\begin{lemma}
    Any distribution that is $\epsilon$-close (in the Wasserstein distance) to the Dirac delta at a given point must have nonzero support within the Euclidean distance sphere centred in that point with radius $\epsilon$.
\end{lemma} 
This means that there is non-zero support on the $\epsilon$ sphere around $c$, and therefore there exists a quantum state $\rho_c$ whose list of expectations is $\epsilon$-close to that point. This yields the following result.
\begin{equation}
    \forall c\in \{-1,1\}^M, \exists \rho_c, \sum_{m=1}^M (\Tr(O_m \rho_c) - c_m)^2 \leq \epsilon^2.
\end{equation}
Because the above is a sum of positive components, we can write $\forall m$
\begin{itemize}
    \item[(a)] if $c_m = 0$, then $-1-\epsilon \leq \Tr(O_m \rho_c) \leq -1 + \epsilon$
    \item[(b)] if $c_m = 1$, then $+1-\epsilon \leq \Tr(O_m \rho_c) \leq +1 + \epsilon$
\end{itemize}
The above yields conditions on the spectrum of $O_m$. We note $\lm$ and $\lM$ respectively the minimum and maximum eigenvalues of $O_m$. We define $\gamma \coloneqq 1 - \epsilon$. We know that $\forall \rho, \Tr(O\rho) \geq \lm$ therefore, $-\gamma \geq \lm$, the same reasoning applies for the maximum eigenvalue, yielding:
\begin{itemize}
    \item[(a)] $\lm \leq -\gamma$,
    \item[(b)] $\lM \geq +\gamma$.
\end{itemize}

For the rest of the proof, we combine approaches from the proof of theorem 2.6 in \cite{aaronson_learnability_2007} and that of B.1 in \cite{jerbi_shadows_2023}. We define the two-outcome POVMs $\{E_m,I-E_m\}$ with
\begin{equation}
    E_m = \frac{O_m - \lm}{\lM-\lm}
\end{equation}
We define $\beta \coloneqq \frac{-\lm}{\lM-\lm}$. The spectral inequalities yield $ \beta \geq \frac{1}{2}$. With this definition, the two conditions above translate to:
\begin{itemize}
    \item[(a)] $c_m = 0, p_0 \coloneqq \Tr(E_m \rho_c) \leq \beta - \frac{\gamma}{\lM-\lm}$
    \item[(b)] $c_m = 1, p_1 \coloneqq \Tr(E_m \rho_c) \geq \beta + \frac{\gamma}{\lM-\lm}$
\end{itemize}

We define the amplified Positive Operator-Valued Measures (POVMs) that apply $\{E_m,I-E_m\}$ to $L \geq 1$ copies of $\rho$ and return 1 if and only if at least $\beta L$ copies of the original POVMs return 1. 

From Holevo's bound (found as Theorem 5.1 in \cite{ambainis_dense_2002}), for the amplified scheme to correctly identify the corresponding bit $b_m = (c_m+1)/2$ with probability $q$ it is necessary that
\begin{equation}
    nL \geq (1-H(q)) M,
\end{equation}
where $H$ is the binary entropy function. 

We define the random variable $X_{m,l}$ which takes the value of the output of the POVM of the $m$-th observable on the $l$-th copy. We define $X_m^{(L)} \coloneqq \frac{1}{L}\sum_l X_{m,l}$.

In the case $c_m=-1$, we have $\mathbb{E}[X_m^{(L)}] = p_0$. The probability of the amplified POVMs yielding the wrong output is
\begin{equation}
    P(X_m^{(L)} > \beta) \leq P\left(X_m^{(L)} > p_0 + \frac{\gamma}{\lM-\lm}\right)
\end{equation}
Recalling that $\beta\geq 1/2$, we can use the Chernoff bound on the Bernoulli variable $X_m^{(L)}$ and we get 
\begin{equation}
    P(X_m^{(L)} > p_0 + \frac{\gamma}{\lM-\lm}) \leq \exp{-\frac{\gamma^2 L }{2 (-\lm) \lM}}
\end{equation}
We define $\Lambda = (-\lm) \lM$, we have $\gamma^2 \leq \Lambda \leq \norm{O}^2$.

For the probability of the amplified POVMs to yield the correct output with probability $q$ it is necessary that $P(X_m^{(L)} \leq \beta) \geq q$.

Finally, we get 
\begin{equation}
    \log(1/q) \leq \frac{\gamma^2 L }{2 \Lambda}
\end{equation}

Combining Chernoff's and Holevo's inequalities, we conclude the proof of \Cref{thm:necessary}:
\begin{equation}
    \Lambda \geq \frac{1-H(q)}{\log(1/q)} \frac{\gamma^2 M}{n}.
\end{equation}

\textbf{Note}: The above is a tighter condition than in Theorem 2.6 in \cite{jerbi_shadows_2023} but falls back to it, when $\Lambda = \norm{O}^2$, which corresponds to $\lm = - \norm{O}$ and $\lM = \norm{O}$. In the opposite scenario, we have $\Lambda = \gamma^2$, which corresponds to constant norm observables, yielding $n \in \Omega(M)$. The norm of observables affects the number of measurements to reach a desired additive accuracy.

\section{Approximation of expectation values}
\label{app:approxexp}
In practice, we do not have access to exact expectation values and we have to estimate them through sampling. This creates a distribution $p_{\hat{Y}}$ ($n$ dimensional) slightly different from the distribution with exact expectation values $p_Y$. For simplicity, we assume that shot noise $\rho_{\epsilon} \sim \mathcal{N}(0,\epsilon I_n)$ ($n$ dimensional) acts like an additional Gaussian noise with zero mean and standard deviation $\epsilon$. $\rho_{\epsilon}$ and $Y$ are independent and we consider the random variable $\hat{Y} = Y + \rho_{\epsilon}$. Therefore the density $p_{\hat{Y}}$ is the convolution of $p_Y$ and $p_{\rho_{\epsilon}}$. Using Lemma 7.1.10 from \cite{ambrosio_gradient_2005}, we know that the Wasserstein distance between $p_Y$ and $p_{\hat{Y}}$, $W_p(p_Y,p_{\hat{Y}}) \in O(\epsilon)$.

In practice, different techniques exist to estimate expectation values of observables with different degrees of measurement efficiency, with shadow tomography techniques \cite{chen_preparation_nodate} surpassing the ``vanilla estimation". For simplicity, we consider a vanilla estimation where each observable with norm $\norm{O}$ is measured $t$ times and the average is returned. This yields a shot noise close to the Gaussian model above with $\epsilon^2 \in \Theta(\norm{O}/ t)$. The total number of measurements $T$ is then $T=tM$, which yield $T \in \Theta(M \norm{O} / \epsilon^2)$

Note that the first result cannot be trivially applied to techniques such as shadow tomography \cite{chen_preparation_nodate}. Indeed the corresponding shot noise $\rho$ cannot in general be modelled by a Gaussian independent noise, at the least the covariance matrix will in general not be proportional to the identity. 
\end{document}